\documentclass[aps,prd,12pt]{revtex4}

\begin{document}
\def\erf{\mbox{erf}}
\def\ER{E_{\rm R}}
\def\eth{\epsilon}
\def\emx{\varepsilon}
\def\vesc{v_{\rm esc}}
\def\SD{{\rm SD}}
\def\SI{{\rm SI}}
\renewcommand{\baselinestretch}{1.0}

\title{One needs positive signatures for detection of Dark Matter}

\author{V.A.~Bednyakov}
\affiliation{Dzhelepov Laboratory of Nuclear Problems,
         Joint Institute for Nuclear Research, \\
         141980 Dubna, Russia; E-mail: Vadim.Bednyakov@jinr.ru}


\begin{abstract}
    One believes there is huge amount of Dark Matter particles in our Galaxy which
    manifest themselves only gravitationally.
    There is a big challenge to prove their existence in a laboratory experiment. 
    To this end it is not sufficient to fight only for the best exclusion curve, 
    one has to see an annual recoil spectrum modulation --- 
    the only available positive direct dark matter detection signature. 
    A necessity to measure the recoil spectra 
    is stressed.

\vskip 0.3cm 
 
 
\end{abstract}

\maketitle

        Galactic Dark Matter (DM) particles do not emit (or reflect) 
	any detectable electromagnetic radiation and 
	manifest themselves only gravitationally by 
	affecting other astrophysical objects.
	According to the estimates based on a detailed model of our Galaxy
\cite{Kamionkowski:1997xg}
 	the local density of DM (nearby the solar system) amounts to about
$\rho_{\rm local}^{\rm DM} \simeq 0.3 \ {\rm GeV/cm}^3
\simeq 5 \cdot 10^{-25} {\rm g/cm}^3$ (see also recent reviews 
\cite{Drees:2012ji,Saab:2012th}). 
	The local flux of DM particles $\chi$ is expected to be 
$\displaystyle \Phi_{\rm local}^{\rm DM} \simeq \frac {100 \ {\rm GeV}} {m_\chi}
\cdot 10^5 \ {\rm cm}^{-2} {\rm s}^{-1},$ where $m_\chi$ is the DM particle mass.
	This value is often considered as a promising basis for 
	direct laboratory dark matter search experiments.
 
	The problem of the DM in the Universe is a challenge for modern physics and 
	experimental technology.
	To solve the problem, i.e. {\em at least}\/ to detect the DM particles, 
	one simultaneously needs to apply the front-end knowledge 
	of modern Particle Physics, Astrophysics, Cosmology and Nuclear Physics 
        and to develop and use over long time 
	extremely high-sensitive experimental setups and
	complex data analysis methods
(see, for example, recent discussion in 
\cite{Cerulli:2012dw}).

	Weakly Interacting Massive Particles (WIMPs) are among the
	most popular candidates for the relic DM. 
	These particles are non-baryonic and there is no room for them in the 
	Standard Model of particle physics (SM).
        The lightest supersymmetric (SUSY) particle (LSP), neutralino
        (being massive, neutral and stable), is currently 
        often assumed to be a favorite WIMP dark matter particle.

        The nuclear recoil energy due to elastic WIMP-nucleus 
        scattering is the main quantity to be measured 
        by a terrestrial detector in direct DM detection laboratory experiments
\cite{Goodman:1985dc}. 
        Detection of the very rare events of such WIMP interactions 
        is a quite complicated task because of
        very weak WIMP coupling with ordinary matter. 
        The rates expected in the SUSY models range from 10 to 10$^{-7}$ 
        events per kilogram detector material a day
\cite{Jungman:1996df,Lewin:1996rx,Ellis:2003eg,Vergados:1996hs,Chattopadhyay:2003xi,%
Bednyakov:2002mb,Bednyakov:1998is,Bednyakov:1997jr}. 
         Moreover, for WIMP masses between a few GeV$/c^2$ and 1 TeV$/c^2$, the energy
         deposited by the recoil nucleus is less than 100 keV.
         Therefore, in order to be able to detect a WIMP, 
         an experiment with a low-energy threshold 
         and an extremely low radioactive background is required. 
         Furthermore, to certainly detect a WIMP one has to 
         unambiguously register some positive signature
         of WIMP-nucleus interactions (directional recoil or
         annual signal modulation)
\cite{Freese:1987wu,Lewin:1996rx}. 
         This means one has to perform a stable measurement with 
	 a detector of large target mass during 3-5 years under
         extremely low radioactive background conditions.
	 There are also some other complications discussed recently in  
\cite{Drees:2012ji,Cerulli:2012dw}. 

        Till now only the DAMA (DArk MAtter) collaboration 
\cite{Bernabei:2003za,Belli:2011kw,Cerulli:2012dw}
        has certainly  observed the first evidence for the DM signal
        due to model-independent registration of the predicted annual
        modulation of specific shape and amplitude due to the combined motions 
	of the Earth and the Sun around the galactic center
\cite{Freese:1987wu}.
        This experiment 
	has released a total exposure of 1.17 t $\times$ yr
	over 13 annual cycles, obtaining  positive 
        model-independent evidence for the presence of DM particles in the galactic 
	halo at 8.9 $\sigma$ C.L.
\cite{Bernabei:2003za,Belli:2011kw,Cerulli:2012dw}.

        Although there are other experiments like
        EDELWEISS, CDMS, XENON, CRESST, etc, which give sensitive exclusion curves,
        no one of them at present has the sensitivity to look for the modulation effect.
        Due to the relatively small target masses and short running times 
	these experiments are unable to see a positive 
        annual modulation signature of the WIMP interactions.
        Unfortunately, some other experiments with 
        targets of much larger mass (mostly NaI) 
        were also unable to register the positive signature due to 
        not good enough background conditions
\cite{Alner:2005kt,Cebrian:2002vd,Yoshida:2000df}.

	Despite the strong and reliable belief of the DAMA collaboration 
	in the observation the annual modulation signature,  
	it is obvious that such a serious claim
	should be verified by at least another one completely independent experiment.  

	If one wants to confirm (more important, if one wants to reject) the DAMA result, 
	one should perform a new experiment which would have 
	the same or better sensitivity to the annual modulation signature
	(and also it would be reasonable  to locate this new 
	setup in another low-background underground laboratory).
	In particular, search 
        for the modulation 
        could be carried out by new-generation experiments
	with high purity germanium detectors of large enough mass, perhaps, 
	both with spin $^{73}$Ge and spinless natural Ge 
\cite{Bednyakov:2008gv}.
        It is interesting that recently the CoGeNT experiment with a 
        germanium detector has reported 
       some preliminary positive indication of the annual modulation
\cite{Aalseth:2011wp}. 

	Together with necessary figthing against backgrounds, 
         the main direction in development of new-generation DM detectors  
	 concerns remarkable enlargement of the target mass to allow observing these
         positive signatures and thus detecting DM and proving or disproving the DAMA claim.
         In particular, an enlarged version of the 
	 EDELWEISS setup with 40 kg bolometric Ge detectors 
\cite{Sanglard:2006hd} together with, perhaps, SuperCDMS 
\cite{Akerib:2006rr,Brink:2005ej}, as well as the enlarged ZEPLIN 
\cite{Akimov:2006qw} or KIMS 
\cite{Lee.:2007qn} experiments might become sensitive to the annual modulation in the future.

\smallskip
       To estimate the expected direct detection rate for these WIMPs
       (in particular, neutralinos) any SUSY-like model
       or some measured data, for example, from the DAMA experiment 
\cite{Bednyakov:2005qd}, can be used. 
       On this basis the WIMP-proton and WIMP-neutron spin $\sigma^{p,n}_{\SD}(0)$  
       and scalar $\sigma^{p,n}_{\SI}(0)$ cross sections at zero-momentum transfer
       can be calculated (see the Appendix). 
       These calculations are usually compared with
       measurements, which (with the only exception of the DAMA result)
       are presented in the form of 
       exclusion curves --- upper limits of the cross section as functions of the WIMP mass.
       In the case of non-observation of any DM signal 
       the exclusion curve simply reflects the sensitivity 
       of a given direct DM search experiment and potentially 
       allows one to constrain some version of the SUSY-like
       theory, if the curve is sensitive enough.
       Therefore the best exclusion curve is currently a clear aim of almost all  
       dark matter search experiments (DAMA/LIBRA and CoGeNT are the only exceptions).
       The main competition between the experiments 
       is in the field of these exclusion curves.

       Before 2000 all exclusion curves
       were evaluated mainly in the one-coupling dominance approach
       (when only one cross section limit was defined from measurements 
       for fixed WIMP mass), 
       which gave slightly pessimistic (for spin-non-zero target
       experiments), but universal limits for all experiments.
       One would say that the competition between the DM experiments was honest. 
       The predictions from SUSY-like models were 
       in general far from being reached by the data. 
       
       Mainly after the paper
\cite{Tovey:2000mm} 
       was published in 2000 (and as well after the DAMA evidence
\cite{Bernabei:2003za}) a new kind of exclusion curves appeared. 
	In particular, for the first time these curves were 
	obtained for the spin-dependent WIMP-nucleon cross section
	limits when non-zero subdominant spin WIMP-nucleon 
	contributions were also taken into account
\cite{Ahmed:2003su,Miuchi:2002zp}.
        This procedure obviously improved the quality of the exclusion curves.
        Therefore a direct comparison of the old-fashioned exclusion curve 
        with the new one could in principle bring one to a wrong
	conclusion about better sensitivity of more recent experiments. 
        There is generally possible incorrectness 
        in the direct comparison of the exclusion curves 
        for the WIMP-proton(neutron) spin-dependent cross section 
        obtained with and without the
        non-zero WIMP-neutron(proton) spin-dependent contribution.
        Furthermore, the above-mentioned incorrectness concerns 
        to a great extent the direct comparison
        of the spin-dependent exclusion curves obtained with and without non-zero
        spin-independent contributions
\cite{Bernabei:2003za,Bernabei:2003wy}.
        Taking into account both spin couplings $a_p$ and $a_n$ but ignoring 
        the scalar coupling $c_0$ (see the Appendix for definitions), 
	one can easily arrive at a misleading conclusion 
        especially for not very light target nuclei when it is not obvious that 
        (both) spin couplings dominate over the scalar one.
        To be consistent,  one has to use the mixed spin-scalar coupling approach
        as was first proposed by the DAMA collaboration
\cite{Bernabei:2000qi,Bernabei:2003za,Bernabei:2003wy}. 
 
         This approach was used in 
\cite{Bednyakov:2008zz} 
         to demonstrate, by the example of the  
         HDMS experiments with natural Ge and with 
         the neutron-odd group high-spin isotope $^{73}$Ge
\cite{Klapdor-Kleingrothaus:2002pg,Klapdor-Kleingrothaus:2000uh}, 
         how one can strongly improve the exclusion curves. 
         The approach allowed both upper limits for the spin-dependent
         $\sigma^{n(p)}_{{\rm SD}}$ and spin-independent 
         $\sigma^{}_{{\rm SI}}$ cross sections of the 
         WIMP-nucleon interaction to be {\em simultaneously}\/ 
         determined from the experimental data.
         In this way visible (one order of magnitude) improvement in the 
         form of the exclusion curves was achieved 
\cite{Bednyakov:2008zz} relative to
         the traditional one-coupling dominance scheme used previously 
	 for the same setup
\cite{Klapdor-Kleingrothaus:2005rn}. 

        As a by-product of the approach, there are correlations 
	(first mentioned in 
\cite{Bernabei:2001ve}) 
        between the mesured upper limits $\sigma^{n}_{{\rm SD}}$ and $\sigma^{}_{{\rm SI}}$, which  
        can be considered as a new requirement --- 
	for any fixed WIMP mass $m_\chi$ one should have \
        $\sigma^{}_\SI({\rm theor.}) \le \sigma^{}_\SI({\rm exp.})$ \ and \
        $\sigma^{n}_\SD({\rm theor.}) \le \sigma^{n}_\SD({\rm exp.})$
        simultaneously, provided that $\sigma^{n(p)}_{\SD(\SI)}({\rm theor.})$
        are calculated in some underlying SUSY-like theory.

\smallskip
	It is important to note 
	that without proper knowledge of the
	nuclear and nucleon structure it is not possible to 
	extract reliable and useful information (at least in the form of these 
	$\sigma^{n}_{{\rm SD}}$ and
        $\sigma^{}_{{\rm SI}}$ cross sections) 
	from direct DM search experiments.
	However,  astrophysical uncertainties, in particular the DM 
	distribution in the vicinity of the Earth
\cite{Copi:2002hm,Kurylov:2003ra,Tucker-Smith:2004jv,
Savage:2004fn,Gondolo:2005hh,Gelmini:2005fb}, make it 
        far more dificult to interpret the results of 
	the DM search experiments.
	At the moment, to have a chance to compare sensitivities
	of different experiments, people adopted a
	common truncated Maxwellian DM particle distribution, 
	but nobody can prove its correctness.
	In the case of undoubted direct DM detection
	one can make some conclusions about the real
	DM particle distribution in the vicinity of the Earth.

       Furthermore, almost by definition (from the very beginning), 
       a modern experiment aiming at the best exclusion curve 
       is doomed to non-observation of the DM signal. 
       This is due to the fact that a typical expected DM signal 
       spectrum exponentially drops
       with recoil energy  and it is practically  
       impossible to single it out from the background non-WIMP
       spectrum of a typical (semiconductor) detector. 

       In fact, one needs a clear, so-called ``positive'' signature
       of interactions between WIMP particles and target nuclei.
       Only exclusion curves are not enough.
       Ideally, this signature should be a unique feature of such an interaction
\cite{Spooner:2007zh}.

       There are some typical characteristics 
       of WIMP particle interactions with a nuclear target which 
       can potentially play the role of these positive WIMP signatures
\cite{Gascon:2005xx}. 
   First of all, WIMPs produce nuclear recoils, 
   whereas most radioactive backgrounds produce electron recoils. 
   Nevertheless, for example, neutrons 
   (and any other heavy neutral particle) can also produce nuclear recoils. 
   There are also proposals 
   which rely on WIMP detection via electron recoils 
\cite{Vergados:2005as,Vergados:2004qj}.

   Due to the extremely rare event rate of the WIMP-nucleus interactions
   (the mean free path of a WIMP in matter is of the order of a light year),
   one can expect two features.
   One is that the probability of two
   consecutive interactions in a single detector or two 
   closely located detectors is completely negligible. 
   Multiple interactions of photons, gamma rays or neutrons under the
   same conditions are much more common. 
   Therefore only non-multiple interaction events 
   can claim  to be from WIMPs.
   The other feature is a uniform distribution of the 
   WIMP-induced events throughout a detector.
   This feature can also be used in the future to identify background 
   events (from photons, neutrons, beta and alpha particles) 
   in rather large-volume position-sensitive detectors.

    The shape of the WIMP-induced recoil energy spectrum
    can be predicted rather accurately (for given WIMP mass, 
    fixed nuclear structure functions, and astrophysical parameters). 
    The observed energy spectrum, claiming to be from WIMPs, 
    must be consistent with the expectation. 
    However, this shape is exponential, right as it is the case for many
    background sources.

   Unfortunately, the nuclear-recoil feature, the non-multiple interaction,
   the uniform event distribution throughout a detector,
   and the shape of the recoil energy spectrum
   could not be the clear ``positive signature'' of the WIMP interactions.  
   It is believed that the following three features of WIMP-nucleus 
   interaction can serve as a clear ``positive signature''.

   The currently most promising, technically feasible 
   and already used (by the DAMA collaboration) ``positive signature'' is
   the annual modulation signature (see the Appendix). 
   The WIMP flux and its average kinetic energy vary annually 
   due to the combined motions of the Earth and the Sun 
   relative to the galactic center.
   The impact WIMP energy increases (decreases) when 
   the Earth velocity is added to (subtracted from) the velocity of the Sun.
      The amplitude of the annual modulation depends on 
      many factors --- 
      details of the halo model, mass of the WIMP, 
      the year-averaged rate (or total WIMP-nucleus cross sections), 
      etc.
      In general, the expected modulation amplitude is rather small
\cite{Freese:1987wu,Lewin:1996rx,Bernabei:2003za,Bernabei:2003wy} 
      and to observe it, one needs huge (at best tonne scale)  detectors 
      which can continuously operate for 5--7 years.
      Of course, to reliably use this signature one should prove 
      the absence of annually modulated backgrounds.

      Another potentially promising positive WIMP signature 
      is connected with the possibility of measuring the 
      direction of the recoil nuclei induced by a WIMP.
      In these directional recoil experiments it is planned
      to measure the correlation of the event rate with the Sun's motion
\cite{Vergados:2003pk,Vergados:2002bb,Vergados:2004qj}.
      Unfortunately, the task is extremely complicated 
\cite{Morgan:2004ys,Sekiya:2004ma,Alner:2004cw,Snowden-Ifft:1999hz,Gaitskell:1996cv}.

        The third well-known potentially useful positive WIMP signature 
	is connected with the coherence of the WIMP-nucleus 
	spin-independent interaction. 
	Due to a rather low momentum transfer, a WIMP 
	coherently scatters by the whole target nucleus and 
	the elastic cross section of this interaction 
	should be proportional to $A^{2}$, where $A$ is 
	the atomic number of the target nucleus.
	Contrary to the $A^2$ behavior, the cross section of 
	neutron scattering by nuclei 
	(due to the strong nature of this interaction)
	is proportional to the geometrical cross-section of the target nucleus 
	($A^{2/3}$ dependence). 
	To reliably use this $A^2$ signature, one has to satisfy
	at least two conditions.
	First, one should be sure that the spin-independent 
	WIMP-nucleus interaction really dominates over
	the relevant spin-dependent interaction. 
	This is far from being obvious
\cite{Bednyakov:2008zz,Bednyakov:2004be,Vergados:2005ky,Vergados:2004hw}.
         Second, one should, at least 
	 for two targets with a different atomic number $A$,   
	 rather accurately 
	 measure the recoil spectra 
	 (in the worst case integrated event rates) 
	 under the same background conditions.
	 Currently, this goal looks far from being achievable. 
	 Developing further the idea of this third signature, 
	 one can also consider as a possible extra WIMP signature
	 an observation of the similarity (or coherent behavior)
	 of measured spectra at different (also non-zero spin) nuclear targets.
	 This possibility relies on rather accurate spin structure functions 
	 for the experimentally interesting nuclei
\cite{Bednyakov:2006ux,Bednyakov:2004xq}. 

        Ideally, in order to be convincing, an eventual DM signal should combine
        more than one of these positive DM signatures
\cite{Gascon:2005xx,Spooner:2007zh}. 

        In the case of currently very promising event-by-event
	active background reduction techniques (like in 
       the CDMS, EDELWEISS and XENON experiments),
       one inevitably needs clear positive WIMP signature(s).
       Without these signatures one can hardly convince anybody that 
       the final spectrum is saturated by WIMPs.
       Furthermore, with the help of these extra signatures and
       on the basis of measured recoil spectra
       one can estimate the WIMP mass 
\cite{Green:2007rb,Shan:2007vn}.

        It is known (see, for example, discussions in 
\cite{Gondolo:2005qp,Bednyakov:1999vh}) 
        that a proof of the observation of a DM signal
        is an extremely complicated problem.
	As pointed out above, on this way an interpretation of measurements  
        in the form of exclusion curves helps almost nothing. 
        Of course, an exclusion curve is at least something from nothing observed.
        It allows a sensitivity comparison
        of different experiments and therefore allows  
        deciding who at the moment is the best 'excluder'.
        But, for example, supersymmetric theory is, 
        in general, very flexible, it has a lot of parameters, 
	and one hardly believes that an exclusion curve 
	can ever impose any decisive constraint on it. 
	Furthremore, 
	almost all experimental groups presenting their 
	exclusion curves try to compare them with some SUSY predictions. 
	It is clear from this comparison 
	that there are some domains of the SUSY parameter
	space, which are now already excluded by these exclusion curves.
	What is remarkable, however, is that nobody yet has seriously 
	considered these constraints for SUSY. 	

        The situation is much worse due to the already mentioned 
        famous nuclear and astrophysical uncertainties involved
        in the evaluation of the exclusion curves 
\cite{Kinkhabwala:1998zj,Donato:1998pc,Evans:2000gr,%
Green:2000jg,Green:2001xy,Copi:2000tv,Ullio:2000bf,Vergados:2000cp}.
        This is why
	it does not look very 
        decisive (or wise) to use very refined 
        data and methods (nuclear, astrophysical, numerical, statistical 
\cite{Feldman:1997qc}, etc)
        and spend big resources fighting 
        only for the best exclusion curve.
	This fighting could only be accepted
	when one tries to strongly improve the sensitivity of
	a small detector with a view 
        of using many copies
	of it in a huge detector array with a total tonne-scale mass
\cite{Bednyakov:2008gv}.

        There are remarks concerning comparison of results from DM search 
	experiments with passive (off-line) background reduction (like DAMA)
        and from experiments with active (on-line) background reduction 
	(like CDMS, XENON, ZEPLIN, etc).
        First, it was demostrated 
\cite{Tovey:2000mm,Ahmed:2003su,Miuchi:2002zp,Bednyakov:2008zz} 
        that any extra positively defined background-like contribution 
        to the spectra improves the extracted (upper limit) values of the cross section.
        Next, within the passive background reduction scheme  
        the measured spectrum is not affected by 
        hardware or software influence during the data taking.
        Further background reduction can be done off-line 
        on the basis of careful investigation of the spectrum itself or, 
        for example, with the help of the pulse shape analysis.
        In this case the extracted background contribution is under control and well defined.
        On the other side, within the active background reduction approach the 
        measured spectrum already contains results
        of this active reduction influence on the data taking process.
        In this case it is not simple to hold under control the real level 
        of extracted on-line background contribution which can easily be overestimated
(see, for example, the relevant discussion in  
\cite{Cerulli:2012dw}). 
        Therefore, due to this obvious difference,
        a direct comparison of exclusion curves from 
        experiments with passive and active background
        reductions could be, in principle, rather misleading.

	Finally, it seems that at the level of our present knowledge 
     the DM problem could not be solved independently 
     of other related problems (proof of SUSY, 
     astrophysical dark matter properties, etc).
     Furthermore, 
     due to the huge complexity of the DM search (technical, physical, astrophysical, 
     necessity for positive signatures, etc),   
     one should deal with the DM problem boldly using 
     a reliable model-dependent framework --- 
     for example the framework of SUSY, where the same LSP neutralino 
     should be seen coherently or lead to effects
     in all available experiments 
     (direct and indirect DM searches, 
     rare decays, high-energy searches at LHC, etc).
     Only if such a SUSY framework leads to a specific and decisive 
     positive WIMP signature, this could mean a proof of SUSY and 
     simultaneous solution of the dark matter problem.
     In some sense, this SUSY framework can serve as a specific and 
     very decisive positive WIMP signature. 


\subsection*{SUMMARY} 

        A physical reason to {\em improve}\/ an exclusion curve is 
         usually an attempt
	to constrain a SUSY-like model. 
	Unfortunately, this is almost hopeless due to the huge flexibility of 
	these models and the inevitable necessity of having extra
	information from other SUSY-sensitive observables (for example, from LHC).
        At the present and foreseeable level of experimental accuracy, 
	simple fighting for the best exclusion curve is almost useless 
	either for real DM detection or for substantial restrictions for SUSY.	

       One should inevitably go beyond an exclusion curve.
       New generations of DM experiments right from their beginning 
       {\em should aim at detection}\/ of the DM particles.
       This will require development of new setups, which will be able to register 
       {\em positive signatures} of the DM particles interactions with nuclear targets.

       One should try to obtain a reliable {\em recoil energy spectrum}.
       First, very accurate off-line investigation of the measured spectrum 
       allows one to single out different non-WIMP background
       sources and to perform controllable background subtractions.  
       Second, the spectrum allows one to look for the annual modulation effect, 
       the only currently  available positive signature 
       of DM particle interactions with terrestrial nuclei.
       This effect is not simply a possibility (among many others) 
       of rejecting background (as claimed again receintly in 
\cite{Saab:2012th}), but it is a unique signature 
        which reflects the inner physical properties of the DM interaction with matter. 
	It is a very decisive and eagerly welcomed feature, 
	which is inevitable for the laboratory proof of the DM exsitence.

\smallskip
        This letter was written in connection with Prof.D.I.Kazakov's 60th birthday and 
	contains updated key messages from the extended review 
	``Direct Search for Dark Matter --- Striking the Balance --- and the Future''
\cite{Bednyakov:2008gv}. 

Preparation of this work was supported by the grant of the Ministry 
of Education and Science of the Russian Federation (contract 12.741.12.0150).

\subsection*{APPENDIX}

\begin{figure}[t!] 
\begin{picture}(100,90) 
\put(-2,0){\includegraphics{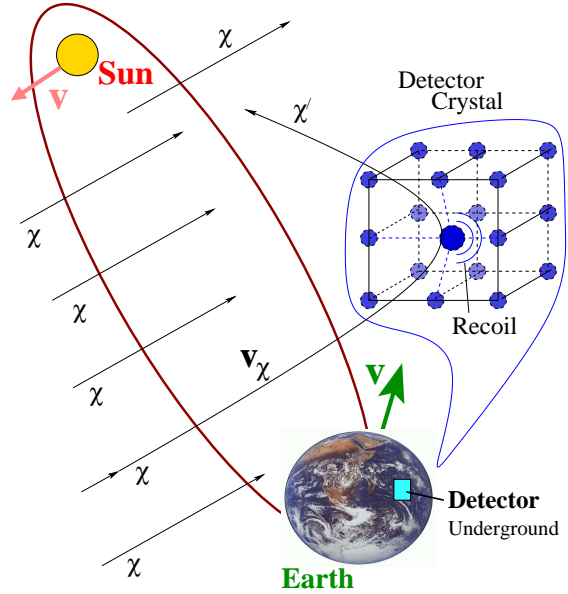}}
\end{picture} 
\caption{Detection of dark matter (WIMPs) by elastic 
  scattering from target nuclei in the detector.
  Due to the expected annual modulation signature of the event rate
  (\ref{Definitions.diff.rate}), the Sun-Earth system 
  is a particularly proper setup for successful direct DM detection.
  From 
\cite{Bednyakov:2008gv}.
\label{DDDDD}}
\end{figure}

	The nuclear recoil energy $E_{\rm R}$ 
	is measured by a proper detector deep underground 
(Fig.~\ref{DDDDD}).
	The differential event rate in respect to the recoil 
	energy (the spectrum) is the subject of the measurements.
	The recoil spectrum produced from WIMP-nucleus scattering in a
	target detector is expected to show the annual
	modulation effect due to the Earth's motion around the Sun
\cite{Freese:1987wu}.
      The velocity of the Earth relative to the
      Galaxy is $v_E(t)=v_S+v_O \cos \gamma \cos \omega(t-t_0)$, where
      $v_S$ is the Sun's velocity relative to the Galaxy ($v_S=232$~km/s), 
      $v_O$ is the Earth's orbital velocity around the Sun 
      ($v_O=30$~km/s) and $\gamma$ is the angle of inclination of the
      plane of the Earth's orbit relative to the galactic plane
      ($\gamma\cong 60^o$). 
      One has $\omega=2\pi/T$ ($T=1$ year) and
      the maximum velocity occurs at day $t_0=155.2$ (June 2).
      The change in the Earth's velocity relative to the incident WIMPs
      leads to a yearly modulation of the scattering event rates of about 7\%. 
      It is convenient to introduce a dimensionless variable $\eta=v_E/v_0$, then
      $ \eta(t)=\eta_0+\triangle\eta \cos{\omega (t-t_0)}$ 
      where the amplitude of the modulated part $(\triangle\eta\simeq 0.07)$ 
      is small compared to the annual average $\eta_0\simeq 1.05$.
      Within this framework, the expected count rate of WIMP interactions
      can be written as 
\begin{equation}
S[\eta(t)] \simeq S[\eta_0]+\left. \frac{\partial S}{\partial \eta}
\right| _{\eta_0}\triangle\eta\cos{\omega (t-t_0)}=S_0+S_m \cos{\omega (t-t_0)}
\end{equation}
     where $S_0$ is the constant part and $S_m$ is the amplitude of the
     modulated signal.
     Both parts of the event rate $S_0$ and $S_m$ depend on the target nucleus $(A,Z)$, 
      WIMP (or neutralino $\chi$) mass ${m_\chi}$,   
     density $\rho_{\rm local}^{\rm DM}$, 
    velocity distribution of the WIMPs in the solar vicinity $f(v)$, 
     and cross section of WIMP-nucleus elastic scattering
(see, for example, 
\cite{Jungman:1996df,Lewin:1996rx,Smith:1990kw,Bednyakov:1999yr}).

	The differential event 
	rate per unit mass of the target material has the form
\begin{equation}
\label{Definitions.diff.rate}
S(t) \equiv	\frac{dR}{dE_{\rm R}} = N_T \frac{\rho_\chi}{m_\chi} 
	\int^{v_{\max}}_{v_{\min}} dv f(v) v
	{\frac{d\sigma^A}{dq^2}} (v, q^2). 
\end{equation}
        Assuming that WIMPs are the dominant component of
        the DM halo of our Galaxy, one has $\rho_{\chi} = \rho_{\rm local}^{\rm DM}$. 
        The nuclear recoil energy
	$E_{\rm R} = q^2 /(2 M_A )$ is typically about $10^{-6} m_{\chi}$, 
	$N_T$ is the number density of target nuclei with mass $M_A$, 
	$v_{\max} = v_{\rm esc} \approx 600$~km/s, and
	$v_{\min}=\left(M_A E_{\rm R}/2 \mu_{A}^2\right)^{1/2}$ is
	the minimal WIMP velocity which still can 
	produce the recoil energy $E_{\rm R}$.
	The WIMP-nucleus differential elastic scattering cross section 
	for spin-non-zero ($J\neq 0$) 
	nuclei contains coherent (spin-independent, or
	SI) and axial (spin-dependent, or SD) terms
\cite{Engel:1992bf,Ressell:1993qm}
\begin{eqnarray} \label{drateEPV}
\frac{d\sigma^A}{dq^2}(v,q^2)
&=&  \frac{S^A_{\rm SD} (q^2)}{v^2 (2J+1)} 
    +\frac{S^A_{\rm SI} (q^2)}{v^2 (2J+1)}
\label{Definitions.cross.section}
= \frac{\sigma^A_{\rm SD}(0)}{4\mu_A^2 v^2}F^2_{\rm SD}(q^2)
           +\frac{\sigma^A_{\rm SI}(0)}{4\mu_A^2 v^2}F^2_{\rm SI}(q^2).
\label{eq:2}
\end{eqnarray}
 	The normalized ($F^2_{\rm SD,SI}(0) = 1$)
	finite-momentum-transfer nuclear form-factors
$\displaystyle
F^2_{\rm SD,SI}(q^2) = \frac{S^{A}_{\rm SD,SI}(q^2)}{S^{A}_{\rm SD,SI}(0)} $
	can be expressed in terms of the nuclear structure functions as follows
\cite{Engel:1992bf,Ressell:1993qm}: 
\begin{eqnarray}
\label{Definitions.scalar.structure.function}
S^{A}_{\rm SI}(q) 
	&=& 
	\sum_{L\, {\rm even}} 
        \vert\langle J \vert\vert {\cal C}_L(q) \vert\vert J \rangle \vert^2 
	\simeq  
	\vert\langle J \vert\vert {\cal C}_0(q) \vert\vert J \rangle \vert^2 ,
\nonumber \\ 
S^A_{\rm SD}(q) 
	&=& 
	\sum_{L\, {\rm odd}} \big( 
	\vert\langle N \vert\vert {\cal T}^{el5}_L(q) 
	\vert\vert N \rangle\vert^2 + \vert\langle N \vert\vert 
	{\cal L}^5_L (q) \vert\vert N \rangle\vert^2\big). 
\label{Definitions.spin.structure.function}
\label{SF-definition}
\label{eq:4}
\end{eqnarray} 
	The explicit form of the transverse electric ${\cal T}^{el5}(q)$ 
	and longitudinal ${\cal L}^5(q)$ multipole projections of the
	axial vector current operator and the scalar function ${\cal C}_L(q)$ 
	can be found in 
\cite{Engel:1992bf,Ressell:1993qm,Bednyakov:2004xq,Bednyakov:2006ux}.
	For $q=0$ the nuclear SD and SI cross sections can be represented as
\begin{eqnarray}
\label{Definitions.scalar.zero.momentum}
\sigma^A_{\rm SI}(0) 
        &=& \frac{4\mu_A^2 \ S^{}_{\rm SI}(0)}{(2J+1)}\! =\!
             \frac{\mu_A^2}{\mu^2_p}A^2 \sigma^{p}_{{\rm SI}}(0), \\ 
\label{Definitions.spin.zero.momentum}
\sigma^A_{\rm SD}(0)
        &=&  \frac{4\mu_A^2 S^{}_{\rm SD}(0)}{(2J+1)}\! =\!
             \frac{4\mu_A^2}{\pi}\frac{(J+1)}{J}
             \left\{a_p\langle {\bf S}^A_p\rangle 
                  + a_n\langle {\bf S}^A_n\rangle\right\}^2\\
\label{Definitions.spin.zero.momentum.Tovei}
&=&
	\frac{\mu_A^2}{\mu_p^2}
	\frac43 \frac{J+1}{J}
	\sigma^{pn}_{\rm SD}(0)	
	\left\{ \langle {\bf S}^A_p\rangle \cos\theta
               +\langle {\bf S}^A_n\rangle \sin\theta   
       \right\}^2.
\label{Definitions.spin.zero.momentum.Bernabei}
\end{eqnarray} 
	Following Bernabei et al.
\cite{Bernabei:2003za,Bernabei:2001ve},
	the effective spin WIMP-nucleon cross section
	$\sigma^{pn}_{\rm SD}(0)$
	and the coupling mixing angle $\theta$ were introduced,
\begin{eqnarray}
\label{effectiveSD-cs}
\sigma^{pn}_{\rm SD}(0)
	&=& \frac{\mu_p^2}{\pi}\frac43 
		\Bigl[ a_p^2 +a_n^2 \Bigr], \qquad
\tan\theta = \frac{{a}_{n}}{{a}_{p}}; \\
\label{effectiveSD-cs-pn} 
\sigma^p_{\rm SD}&=&\sigma^{pn}_{\rm SD} \cdot \cos^2 \theta, \quad
\sigma^n_{\rm SD}=\sigma^{pn}_{\rm SD} \cdot \sin^2 \theta.
\end{eqnarray}
	Here, $\displaystyle \mu_A = \frac{m_\chi M_A}{m_\chi+ M_A}$
	is the reduced mass of the neutralino and the nucleus, 
	and it is assumed that $\mu^2_{n}=\mu^2_{p}$.
	The dependence on effective WIMP-quark (in SUSY neutralino-quark) couplings 
	${\cal C}_{q}$ and ${\cal A}_{q}$ in the underlying theory 
\begin{equation}
{\cal  L}_{\rm eff} = \sum_{q}^{}\left( 
	{\cal A}_{q}\cdot
      \bar\chi\gamma_\mu\gamma_5\chi\cdot
                \bar q\gamma^\mu\gamma_5 q + 
	{\cal C}_{q}\cdot\bar\chi\chi\cdot\bar q q
	\right)      \ + ... 
\label{Definitions.effective.lagrangian}
\end{equation}
	and on the spin ($\Delta^{(p,n)}_q$)
	and the mass or scalar ($f^{(p)}_q \approx f^{(n)}_q$) 
	structure of the proton and neutron 
	enter into these formulas via the zero-momentum-transfer 
	WIMP-proton and WIMP-neutron SI and SD cross sections 
\begin{eqnarray}
\label{Definitions.scalar.zero.cs}
\label{SI-cs-Deff.at-zero}
\sigma^{p}_{{\rm SI}}(0) 
	= 4 \frac{\mu_p^2}{\pi}c_{0}^2,
&\qquad&
	c^{}_0 = c^{p,n}_0 = \sum_q {\cal C}_{q} f^{(p,n)}_q; \\
\sigma^{p,n}_{{\rm SD}}(0)  
 	=  12 \frac{\mu_{p,n}^2}{\pi}{a}^2_{p,n} 
&\qquad&
	a_p =\sum_q {\cal A}_{q} \Delta^{(p)}_q, \quad 
	a_n =\sum_q {\cal A}_{q} \Delta^{(n)}_q.
\label{SD-cs-Deff.at-zero}
\end{eqnarray}
	The factors $\Delta_{q}^{(p,n)}$, which parameterize the quark 
	spin content of the nucleon, are defined as
	$ \displaystyle 2 \Delta_q^{(n,p)} s^\mu  \equiv 
          \langle p,s| \bar{\psi}_q\gamma^\mu \gamma_5 \psi_q    
          |p,s \rangle_{(p,n)}$.
	The quantity $\langle {\bf S}^A_{p(n)} \rangle $ denotes 
	the total spin of protons 
	(neutrons) averaged over all $A$ nucleons of the nucleus $(A,Z)$
\begin{equation}
\langle {\bf S}^A_{p(n)} \rangle 
     \equiv \langle A \vert  {\bf S}^A_{p(n)} \vert A \rangle 
     = \langle A \vert  \sum_i^A {\bf s}^i_{p(n)} \vert A \rangle. 
\end{equation}
        The mean velocity $\langle v \rangle$ of 
	the relic neutralinos of our Galaxy
	is about  $ 300~{\rm km/s} = 10^{-3} c$.
	Assuming $q_{\rm max}R \ll 1$, where $R$ is the nuclear radius 
	and $q_{\rm max} = 2 \mu_A v$ is the maximum of the momentum 
	transfer in the process of the $\chi A$ scattering, the 
        spin-dependent matrix element  takes a simple form
({\em zero momentum transfer limit})\
\cite{Engel:1995gw,Ressell:1997kx}
\begin{equation}
\label{Definitions.matrix.element}
 {\cal M} = C \langle A\vert a_p {\bf S}_p + a_n {\bf S}_n
 	\vert A \rangle \cdot {\bf s}_{\chi}
 	  = C \Lambda \langle A\vert {\bf J}
	 \vert A \rangle \cdot {\bf s}_{\chi}.
\label{eq:10}
\end{equation}
	Here, ${\bf s}_{\chi}$ denotes the spin of the neutralino, and 
\begin{equation}
 \Lambda = {{\langle N\vert a_p {\bf S}_p + a_n {\bf S}_n
\vert N \rangle}\over{\langle N\vert {\bf J}
\vert N \rangle}} =
{{\langle N\vert ( a_p {\bf S}_p + a_n {\bf S}_n ) \cdot {\bf J}
\vert N \rangle}\over{ J(J+1)
}} = 
{{a_p \langle {\bf S}_p \rangle}\over{J}}  +
{{a_n \langle {\bf S}_n \rangle}\over{J}}.
\label{eq:11}
\end{equation}
	The normalization factor $C$ involves the coupling
	constants, the  masses of the exchanged bosons, and 
        the mixing parameters relevant to the LSP, i.e., it is not
        related to the associated nuclear matrix elements
\cite{Griest:1988ma}.  
	In the limit of zero momentum transfer $q=0$ 
	the spin structure function in 
(\ref{Definitions.spin.structure.function}) reduces to the form
\begin{eqnarray*}
S^A(0)={1\over{4 \pi}} \vert\langle A \vert\vert\sum_i
    {1\over{2}}(a_0 + a_1 \tau_3^i) {\bf \sigma}_i\vert\vert A \rangle\vert^2 
     =\frac{2J+1}{\pi} J(J+1) \Lambda^2. 
\end{eqnarray*}
        For the most interesting
	isotopes either $\langle{\bf S}^A_{p}\rangle$ 
	or $\langle{\bf S}^A_{n}\rangle$ dominates
	($\langle{\bf S}^A_{n(p)}\rangle \ll \langle{\bf S}^A_{p(n)}\rangle$).

           The differential event rate 
(\ref{Definitions.diff.rate}) can be also given also in the form 
\cite{Bernabei:2003za,Bednyakov:2004be}
\begin{eqnarray}
\label{Definitions.diff.rate1}
\frac{dR(\ER)}{d\ER} 
	&=& \kappa^{}_\SI(\ER,m_\chi)\,\sigma_\SI
         +\kappa^{}_\SD(\ER,m_\chi)\,\sigma_\SD. \\
\nonumber
\kappa^{}_\SI(\ER,m_\chi)
&=&       N_T \frac{\rho_\chi M_A}{2 m_\chi \mu_p^2 } 
          B_\SI(\ER) \left[ M_A^2 \right],\\
\kappa^{}_\SD(\ER,m_\chi)
&=& 
\label{structure} 
     N_T \frac{\rho_\chi M_A}{2 m_\chi \mu_p^2 } B_\SD(\ER) 
        \left[\frac43 \frac{J+1}{J}
        \left(\langle {\bf S}_p \rangle \cos\theta
            + \langle {\bf S}_n \rangle \sin\theta   
       \right)^2\right]  ,\\
B_{\SI,\SD}(\ER) 
&=& \nonumber
        \frac{\langle v \rangle}{\langle v^2 \rangle}
        F^2_{\SI,\SD}(\ER)I(\ER). 
\end{eqnarray}
        The dimensionless integral $I(\ER)$ 
        is a dark-matter-particle velocity distribution correction 
\begin{equation}\label{I_ER}
I(\ER)= \frac{ \langle v^2 \rangle}{ \langle v \rangle }
 \int_{x_{\min}} \frac{f(x)}{v} dx 
    = \frac{\sqrt{\pi}}{2}
\frac{3 + 2 \eta^2}{{\sqrt{\pi}}(1+2\eta^2)\erf(\eta) + 2\eta e^{-\eta^2}}
                [\erf(x_{\min}+\eta) - \erf(x_{\min}-\eta)],
\end{equation}
        where  WIMPs in the rest frame of our Galaxy are assumed to  
have a Maxwell-Boltzmann velocity distribution,
the dimensionless Earth speed with respect to the halo 
	$\eta$, is used, and 
        $\displaystyle x_{\min}^2 = 
        \frac{3}{4}\frac{M_A\ER}{\mu^2_A{\bar{v}}^2}$
\cite{Freese:1987wu,Lewin:1996rx}. 
        The error function is 
        $\displaystyle \erf(x) = \frac{2}{\sqrt{\pi}}\int_0^x dt e^{-t^2}$.
        The velocity variable is the dispersion $\bar{v}\simeq 270\,$km$/$c.
        The mean WIMP velocity 
        ${\langle v \rangle} = \sqrt{\frac{5}{3}} \bar{v}$.
	Integrating the differential rate 
(\ref{Definitions.diff.rate}) 
	from the recoil energy threshold $\eth$ 
	to some maximal energy $\emx$,
        one obtains the total detection rate $R(\eth, \emx)$
	as a sum of the SD and SI terms
\begin{eqnarray}\label{Definitions.total.rate}
\label{for-toy-mixig}
R(\eth, \emx)&=&
     R_{\SI}(\eth, \emx) + R_{\SD}(\eth, \emx) = 
      \int^{\emx}_{\eth} d\ER \kappa^{}_\SI(\ER,m_\chi)\,\sigma_\SI
    + \int^{\emx}_{\eth} d\ER \kappa^{}_\SD(\ER,m_\chi)\,\sigma_\SD.
\end{eqnarray}
	To accurately estimate the event rate $R(\eth, \emx)$,
	one needs to know a number of quite uncertain 
	astrophysical and nuclear structure parameters
	as well as the very specific characteristics of the experimental setup
\cite{Cerulli:2012dw}.

	As $m_{\chi}$ increases, 
	the product $qR$ becomes non-negligible
	and {\em the finite momentum transfer limit}\/
	must be considered
\cite{Engel:1992bf,Ressell:1993qm,Ressell:1997kx,%
Bednyakov:2004xq,Bednyakov:2006ux}. 
	With the isoscalar spin coupling constant $a_0 = a_n + a_p$
	and the isovector spin coupling constant
	$a_1 = a_p - a_n$, one can split 
	the nuclear structure function $S^A_{}(q)$ 
	into a pure isoscalar term, $S^A_{00}(q)$, a pure isovector term, $S^A_{11}(q)$, 
	and an interference term, $S^A_{01}(q)$, in the following way:
\begin{equation}
\label{Definitions.spin.decomposition}
S^A_{}(q) = a_0^2 S^A_{00}(q) + a_1^2 S^A_{11}(q) + a_0 a_1 S^A_{01}(q).
\end{equation}
	The relations 
$S^A_{00}(0) = C(J)(\langle {\bf S}_p \rangle + \langle {\bf S}_n \rangle)^2,$ 
$S^A_{11}(0) = C(J)(\langle {\bf S}_p \rangle - \langle {\bf S}_n \rangle)^2,$ and 
$S^A_{01}(0) =2C(J)(\langle {\bf S}^2_p \rangle - \langle {\bf S}^2_n \rangle)
$ with
$\displaystyle C(J)= \frac{2J+1}{4\pi}\frac{J+1}{J}$
       connect the nuclear spin structure function $S^A(q=0)$ 
       with the proton $\langle {\bf S}_p \rangle$ 
       and neutron $\langle {\bf S}_n \rangle$
       spin contributions averaged over the nucleus
\cite{Bednyakov:2006ux}. 

        To analysis modern 
	data 
        in the finite momentum transfer approximation 
	it seems reasonable to use the formulas for the  
 	differential event rate 
(\ref{Definitions.diff.rate}) as schematically given below
\begin{eqnarray}
\label{fit-finite-q}
\label{Definitions.spin.differential.rate}
\frac{dR(\eth,\emx)}{d\ER}
&=& 
{\cal N}(\eth,\emx,\ER,m_\chi)
     \left[ \eta^{}_\SI(\ER,m_\chi) \,\sigma^{p}_\SI
           +\eta^{\prime}_\SD(\ER,m_\chi,\omega) \,  {a_0^2}
     \right ];
\\ \nonumber
{\cal N}(\eth,\emx,\ER,m_\chi) 
&=&  \left[ N_T \frac{c \rho_\chi}{2 m_\chi} 
      \frac{M_A}{\mu_p^2}
      \right]
      \frac{4 \mu_A^2}{\left< q^2_{\max}\right>}
      \langle \frac{v}{c} \rangle  I(\ER)
       \theta (\ER-\eth) \theta(\emx-\ER),
\\ \nonumber
\eta^{}_\SI(\ER,m_\chi)
&=& \left\{ A^2 F^2_\SI(\ER)\right\}
; \\ \nonumber
\eta^{\prime}_\SD(\ER,m_\chi,\omega)
&=& \mu_p^2
\left\{ \frac{4 
}{2J+1} 
	\left(S_{00}(q) + \omega^2\, S_{11}(q) + \omega\, S_{01}(q) 
\right) \right\} .
\end{eqnarray}
       Here the 
       isovector-to-isoscalar nucleon coupling ratio 
       is $\omega = {a_1}/{a_0}$.
       The detector threshold recoil energy 
       $\eth$ and the maximal available recoil energy $\emx$ 
       ($\eth \le \ER \le \emx$) have been introduced in 
(\ref{Definitions.total.rate}).
       In practice, with an ionization or scintillation signal, one has to 
       take into account the quenching of the recoil energy, when 
       the visible recoil energy is smaller than the real
       recoil energy transmitted by the WIMP to the target nucleus.

       Formulas 
(\ref{fit-finite-q}) allow experimental recoil 
       spectra to be directly described in terms of only {\em three}\ 
\cite{Bednyakov:1994te} 
       (it is rather reasonable to assume 
       $\sigma^{p}_\SI(0)\approx \sigma^{n}_\SI(0)$) 
       independent parameters 
       ($\sigma^{p}_\SI$, $a^2_0$ and $\omega$)
       for any fixed WIMP mass $m_\chi$ and any neutralino composition.
	Comparing this formula with the observed recoil spectra 
	for different targets (Ge, Xe, F, NaI, etc)
	one can directly and simultaneously 
	restrict both isoscalar $c_0$ (via $\sigma^{p}_\SI$) and isovector 
	neutralino-nucleon effective couplings $a_{0,1}$.
	These constraints,
	based on the nuclear spin structure functions for finite $q$, 
	will impose {\em the most 
	model-independent and most accurate restrictions}\ 
	on any SUSY parameter space.
       Contrary to some other possibilities (see, for example, 
\cite{Bernabei:2003za,Tovey:2000mm}), 
       this procedure is direct and uses
       as much as possible the results of the 
       accurate nuclear spin structure calculations.

       It is seen from (\ref{effectiveSD-cs-pn}) and (\ref{fit-finite-q}) that
        the SD cross sections $\sigma^{p}_{\SD}$ and $\sigma^{n}_{\SD}$ 
	(or equivalently $a^2_{0}$ and $\omega = {a_1}/{a_0}$) are 
        the only two WIMP-nucleon spin variables which can be  
	constrained (or extracted) from DM measurements. 
        Therefore, there is no sense in extracting 
        effective WIMP-nucleon couplings $a^{}_{p}$ and $a^{}_{n}$ 
from the data 
        (with ``artificial'' twofold ambiguity).

        Finally, to estimate the expected direct DM detection rates (with formulas
(\ref{Definitions.diff.rate}),
(\ref{for-toy-mixig}) or (\ref{fit-finite-q}))
        one should calculate the cross sections $\sigma^{}_\SI$ and $\sigma^{}_\SD$ 
	(or WIMP-nucleon couplings $c_0$ and $a^{}_{p,n}$)
	within a SUSY-based model or take them 
	from experimental data (if it is possible).

{\small 
\providecommand{\href}[2]{#2}\begingroup\raggedright\endgroup
 } \end{document}